\documentclass[aps,prl,superscriptaddress,twocolumn,nopacs,amsmath,amssymb,letter,graphicx]{revtex4-1}

\usepackage{color}
\usepackage{graphicx}
\usepackage{dcolumn} 
\usepackage{bm}
\usepackage{xspace}
\setcitestyle{numbers,square}

\begin{document}

\title{A general route to form topologically-protected surface and bulk Dirac fermions along high-symmetry lines}

\author{O.~J.~Clark}
\email{ojc3@st-andrews.ac.uk}
\affiliation {SUPA, School of Physics and Astronomy, University of St. Andrews, St. Andrews KY16 9SS, United Kingdom}

\author{F.~Mazzola}
\affiliation {SUPA, School of Physics and Astronomy, University of St. Andrews, St. Andrews KY16 9SS, United Kingdom}

\author{I. Markovi\'c}
\affiliation {SUPA, School of Physics and Astronomy, University of St. Andrews, St. Andrews KY16 9SS, United Kingdom}
\affiliation {Max Planck Institute for Chemical Physics of Solids, N{\"o}thnitzer Stra{\ss}e 40, 01187 Dresden, Germany}

\author{J.~M.~Riley}
\affiliation {SUPA, School of Physics and Astronomy, University of St. Andrews, St. Andrews KY16 9SS, United Kingdom}
\affiliation{Diamond Light Source, Harwell Campus, Didcot, OX11 0DE, United Kingdom}

\author{J.~Feng}
\affiliation {SUPA, School of Physics and Astronomy, University of St. Andrews, St. Andrews KY16 9SS, United Kingdom}
\affiliation {Suzhou Institute of Nano-Tech. and Nanobionics (SINANO), CAS, 398 Ruoshui Road, SEID, SIP, Suzhou, 215123, China}

\author{B.-J.~Yang}
\affiliation{Department of Physics and Astronomy, Seoul National University, Seoul 08826, Korea}
\affiliation{Center for Correlated Electron Systems, Institute for Basic Science (IBS), Seoul 08826, Korea}
\affiliation{Center for Theoretical Physics (CTP), Seoul National University, Seoul 08826, Korea}

\author{K.~Sumida}
\affiliation{Graduate School of Science, Hiroshima University, 1-3-1 Kagamiyama, Higashi-Hiroshima, 739-8526 Japan}

\author{T.~Okuda}
\affiliation{Hiroshima Synchrotron Radiation Center, Hiroshima University, 2-313 Kagamiyama, Higashi-Hiroshima, 739-0046 Japan }

\author{J.~Fujii}
\author{I.~Vobornik}
\affiliation{Istituto Officina dei Materiali (IOM)-CNR, Laboratorio TASC,
in Area Science Park, S.S.14, Km 163.5, I-34149 Trieste, Italy}

\author{T.~K.~Kim}
\affiliation{Diamond Light Source, Harwell Campus, Didcot, OX11 0DE, United Kingdom}

\author{K.~Okawa}
\affiliation {Laboratory for Materials and Structures, Tokyo Institute of Technology, Kanagawa 226-8503, Japan}

\author{T.~Sasagawa}
\affiliation {Laboratory for Materials and Structures, Tokyo Institute of Technology, Kanagawa 226-8503, Japan}

\author{M.~S.~Bahramy}
\email{bahramy@ap.t.u-tokyo.ac.jp}
\affiliation{Quantum-Phase Electronics Center and Department of Applied Physics, The University of Tokyo, Tokyo 113-8656, Japan}
\affiliation{RIKEN center for Emergent Matter Science (CEMS), Wako 351-0198, Japan} 

\author{P.~D.~C.~King}
\email{philip.king@st-andrews.ac.uk}
\affiliation {SUPA, School of Physics and Astronomy, University of St. Andrews, St. Andrews KY16 9SS, United Kingdom}

\begin{abstract}
{
\noindent The band inversions that generate the topologically non-trivial band gaps of topological insulators and the isolated Dirac touching points of three-dimensional Dirac semimetals generally arise from the crossings of electronic states derived from different orbital manifolds. Recently, the concept of single orbital-manifold band inversions occurring along high-symmetry lines has been demonstrated, stabilising multiple bulk and surface Dirac fermions. Here, we discuss the underlying ingredients necessary to achieve such phases, and discuss their existence within the family of transition metal dichalcogenides. We show how their three-dimensional band structures naturally produce only small $k_z$ projected band gaps, and demonstrate how these play a significant role in shaping the surface electronic structure of these materials. We demonstrate, through spin- and angle-resolved photoemission and density functional theory calculations, how the surface electronic structures of the group-X TMDs PtSe$_2$ and PdTe$_2$ are host to up to five distinct surface states, each with complex band dispersions and spin textures. Finally, we discuss how the origin of several recently-realised instances of topological phenomena in systems outside of the TMDs, including the iron-based superconductors, can be understood as a consequence of the same underlying mechanism driving $k_z$-mediated band inversions in the TMDs. }
\end{abstract}

\maketitle    

\section{Introduction}

Materials hosting Dirac cones in their electronic structures are desirable for their high carrier mobilities~\cite{qu_quantum_2010, chen_intrinsic_2008, liang_ultrahigh_2015}, potential to realise anomalous quantum hall effects~\cite{qu_quantum_2010, gusynin_unconventional_2005} and for the broader study of massless Dirac fermions in laboratory environments. Bulk Dirac cones originate from lattice symmetry-protected crossing points of the bulk electronic structure~\cite{yang_classification_2014}, while parity-inverted bulk band gaps yield spin-polarised Dirac states at the surface, termed topological surface states (TSSs). By now, numerous examples of systems hosting bulk Dirac points (BDPs) or TSSs have been discovered. In almost all of these cases, however, the band inversions which stabilise these phases result from the crossing of electronic states derived from different atomic manifolds. For example, the  prototypical topological insulator, Bi$_2$Se$_3$, has an inverted band gap formed between predominantly $p_z$-derived bands originating from Se and Bi, respectively. The band inversion is evident in the crystal field levels, which invert their ordering as compared to the atomic limit when spin-orbit coupling (SOC) is included. Such band inversions necessarily do not survive a reduction of SOC strength, as evidenced, for example, by the topologically trivial character of the sister compound Sb$_2$Se$_3$~\cite{zhang_topological_2009}. 

In contrast, recent studies of the transition metal dichalcogenides (TMDs) have shown how these compounds generically host topological phenomena that are driven not by the intersection of states derived from different atomic multiplets, but instead from a single orbital manifold, where the band crossings arise due to the disparate bandwidth of $p$-derived bands along the $k_z$ axis of their Brillouin zone (BZ)~\cite{bahramy_ubiquitous_2018}.  Eight TMDs have so far been shown to host at least a bulk Dirac point (BDP) (most often of type-II character) and two topological surface states (TSSs), collectively forming a `topological ladder' centred at the $\overline{\Gamma}$ point of the surface Brillouin zone~\cite{bahramy_ubiquitous_2018, clark_fermiology_2018, yan_identification_2015, noh_experimental_2017, huang_type_2016, zhang_experimental_2017, yan_lorentz_2017, xu_topological_2018, fei_approaching_2017}. Arising from band inversions along a high-symmetry line, rather than at an isolated $k$-point, provides them with an inherent robustness against modest changes to the details of an underlying lattice. For example, SOC sets only the relative starting energy scales of the dispersing bands, and so a reduction in the spin-orbit coupling strength does not destroy either the Dirac cones or, more remarkably, the topological surface states, but instead simply modifies their position along the rotationally symmetric axis~\cite{bahramy_ubiquitous_2018}. 

Here, we will review the fundamental mechanisms driving the formation of $k_z$-mediated topological ladders in the context of the 1T-structured transition metal dichalcogenides (TMDs). Additionally, we will show that the broad bandwidths and high $k_z$ dispersions associated with these systems naturally results in small $k_z$-projected band gaps in which the surface electronic states must reside. We show how this significantly modifies the topologically non-trivial surface states, leading to bands which are far removed from the prototypical conic dispersions and in-plane spin-momentum locked spin textures known from model systems such as Bi$_2$Se$_3$. We conclude by discussing the applicability of the underlying mechanism to other material systems. In particular, we will show that an analogous mechanism drives the formation of many recently realised  instances of bulk Dirac points and topological surface states in compounds ranging from transition metal carbides, to tetragonal Fe-based superconductors.

\section{Methods}
\textbf{(Spin-)ARPES:}
High-quality single crystal samples of PdTe$_2$ and PtSe$_2$, grown by chemical vapour transport, were cleaved \textit{in situ} at measurement temperatures below 22~K.
Spin-integrated ARPES measurements  were performed at the I05 beamline of Diamond Light Source, UK using $p$-polarised photons at energies between $h\nu=$24 and 107~eV~\cite{hoesch_facility_2017}.  Spin-resolved ARPES data in Fig.~\ref{f:PTSpin}(b) and Fig.~\ref{f:PSSpin}(a,b,d) were obtained from beamline BL9A of HiSOR, Japan~\cite{okuda_efficient_2011, okuda_double_2015}. Spin-resolved data in Fig.~\ref{f:PTSpin}(c) was obtained from the APE beamline of Elettra, Italy~\cite{bigi_very_2017}.  In both cases, the finite spin-detection
efficiency was corrected using detector-dependent effective Sherman functions, as determined by fitting the spin-polarisation of reference
measurements of the Bi(111) (BL9A) or Au(111) (APE) Rashba-split surface states.

Spin-resolved energy distribution curves (EDCs) were determined according to

\begin{equation}
    I_i^{\uparrow,\downarrow} = \frac{I_i^{\text{tot}}(1\pm P_i)}{2},
\end{equation}

where $i \in \{x,y,z\}$, $I_i^{\text{tot}}=(I_i^+ + I_i^-)$ and $I_i^{\pm}$ is the measured intensity for a positively or negatively magnetised detector. The final spin polarisation, $P_i$, is defined as follows

\begin{equation}
    P_i = \frac{(I_i^+ - I_i^-)}{SI_i^{\text{tot}}},
\end{equation}
where $S$ is the relevant effective Sherman function.

Spin-resolved dispersions shown in Fig.~\ref{f:PTSpin}(b) and Fig.~\ref{f:PSSpin}(a) utilise a 2D colour scale: The total intensity, $I_i^{\text{tot}}$ is included as a transparency filter over the determined spin-polarisation, $P_i$, to suppress spurious noise arising from low total spectral weight in the background.

\textbf{Calculations:}
The bulk calculations were performed within density functional theory (DFT) using the Perdew-Burke-Ernzerhof exchange-correlation functional as implemented in
the~{\sc wien2k} package~\cite{wien}. Relativistic effects including spin-orbit coupling were fully taken into account. For all atoms, the muffin-tin radius $R_{MT}$  was chosen such that its product with the maximum modulus of reciprocal vectors, $K_{max}$ become $R_{MT}K_{max} = 7.0$. The Brillouin zone sampling of structures was carried out using a 20 $\times$ 20 $\times$ 20 $k$-mesh. For the surface calculations, a 100
unit tight binding supercell was constructed using maximally localized Wannier functions~\cite{souza, mostofi, kunes}. The $p$-orbitals of the chalcogen and the $d$-orbitals of the transition metal atoms were chosen as the projection centres.

\section{$k_z$-mediated band inversions}
\begin{figure*} 
	\includegraphics[width=\textwidth]{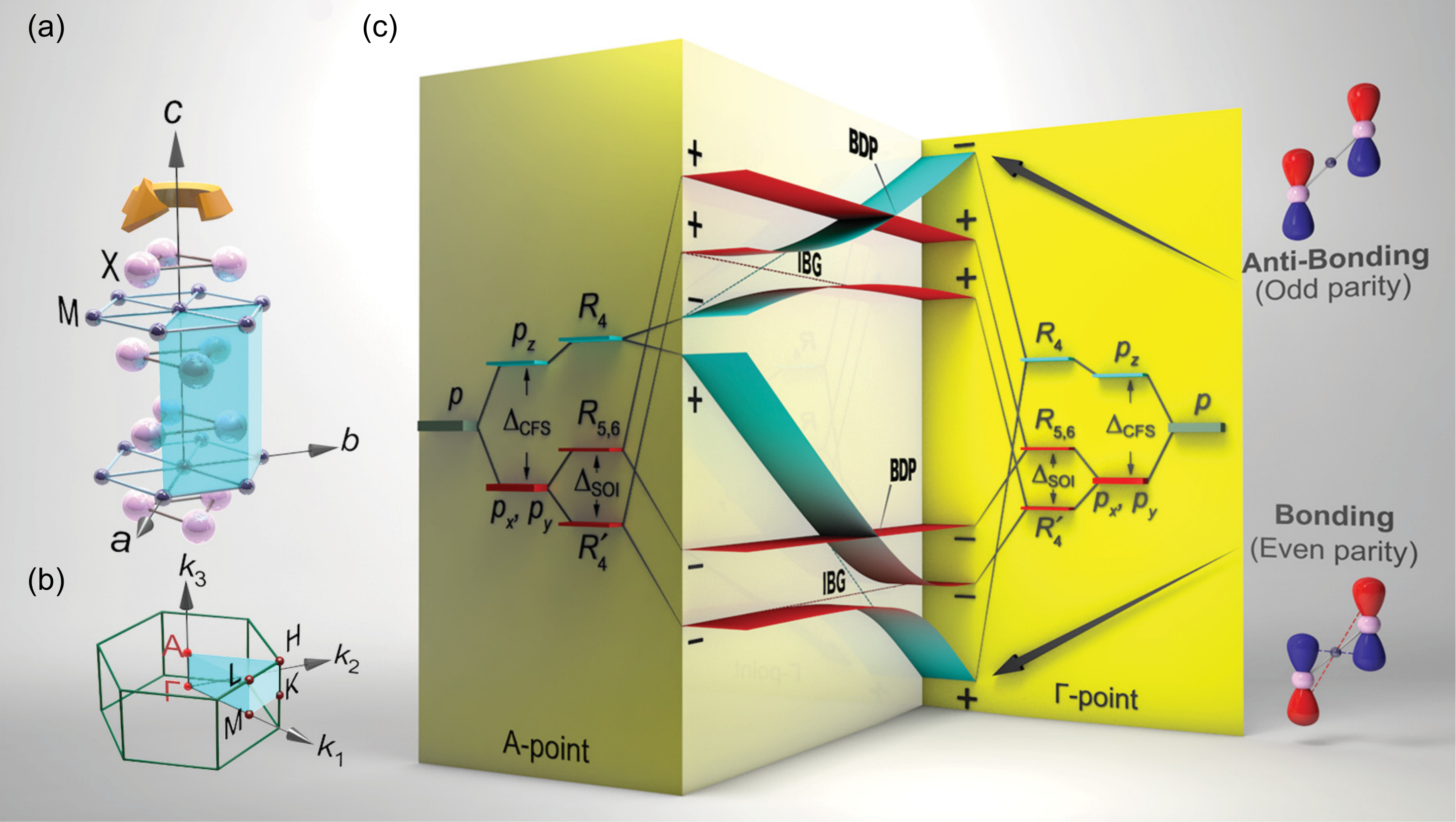}
	\caption{Hierarchy of topologically non-trivial bulk and surface Dirac fermions in transition metal dichalcogenides MX$_2$. (a) Typical crystal structure of the $1T$-type MX$_2$ compounds, composed of a transition metal M at the origin, six-fold coordinated to chalcogen atoms X. The resulting unit cell is three-fold rotationally symmetric along $c$-axis, enforcing the inversion symmetry between the X sites about the M centre. (b) The first Brillouin zone of 1T-MX$_2$ and its corresponding high-symmetry $k$-points. (c) Schematic diagram of $k_z$-mediated topological ladders, stemming from the chalcogen $p$-orbital manifold. The combination of crystal field splitting (CFS), spin orbit interaction (SOI) and bonding-anti-bonding splitting leads to multiple band crossings along $\Gamma$-A direction, some protected by rotational symmetry forming a bulk Dirac point (BDP) and some gapped by SOI, resulting in a local band gap with inverted parity (IBG). The origin of the parity switch for the bonding and anti-bonding $p_z$ states at $\Gamma$ is shown schematically on the right. Similar arguments apply for the other states.}
	\label{f:mechanism}
\end{figure*}

The mechanism underpinning the formation of topological ladders in the TMDs is illustrated in Fig.~\ref{f:mechanism}, in the context of the 1T-structured ($D_{3d}$) compounds. The unit cell of a 1T-TMD (MX$_6$ octahedron) is composed of three sublayers (see Fig.~\ref{f:mechanism}(a)). The central sub-layer is populated by the transition metal, M, positioned at the inversion centre of the unit cell ($\boldsymbol{r}=0$). The six chalcogen atoms of an MX$_6$ octahedron are divided equally between the two sublayers either side of the transition metal, each possessing trigonal symmetry. There is a relative 180 degree rotation between these two sub-layers, such that each chalcogen atom can be mapped onto an inequivalent chalcogen atom by a translation $-\boldsymbol{r} \rightarrow \boldsymbol{r}$, passing through the inversion centre at the transition metal site. In what follows, the role of the transition metal will be neglected to focus entirely on the chalocgen $p$-orbital manifold. This is a reasonable approximation for the group-X systems where the transition metal weight is far removed from the Fermi level, although the arguments presented below are not significantly altered even when there is more significant $d$-orbital mixing, as we return to below.

Fig.~\ref{f:mechanism}(c) shows the energetic hierarchy of the $p$-orbital-derived energy levels at both the $\Gamma$ ($\mathbf{k}=(0,0,0)$) and A ($\mathbf{k}=(0,0,\pi/c)$) high-symmetry points of the Brillouin zone, for the relevant $D_{3d}$ octahedral symmetry. Starting from triply degenerate (neglecting spin) $p_{x,y,z}$ orbitals, the crystal field splitting (CFS) acts to separate them into an $A_1$ ($p_z$) and $E$ ($p_{x,y}$) symmetry manifolds. The spin-orbit interaction (SOI) modifies the energetic separation of $A_1$ and $E$ as well as further splitting the latter into the $p_{x,y}$-derived $R_{5,6}$ and $R_{4}'$, where $R_i$ denotes a double group representation~\cite{doublegroup}, with spin-orbit coupling included. The $p_z$-derived band has $R_{4}$ symmetry. 

Critically, there are two chalcogen sites within the unit cell. Bonding (B) and anti-bonding (AB) combinations of these are thus created. 
At the $\Gamma$-point, the isophasic interference of chalcogen wave functions $\psi_{\text{X}}(\boldsymbol{r})$ leads to a large B-AB splitting of the $p_z$-derived states. The B-AB splitting is still large for the $p_{x,y}$-derived bands, but reduced as compared to the $p_z$ states due to the frustration of the trigonal lattice for the in-plane $p_{x,y}$ bands. At the A-point, the acquired phase factor $e^{i\boldsymbol{k}.\boldsymbol{r}}$ enforces a destructive interference between $\psi_{\text{X}}(\boldsymbol{r})$ of the neighbouring chalcogen sublayers, thereby resulting  in a significant reduction of B-AB splitting of the $p_z$ states, and accordingly a large $k_z$-dispersion. For simplicity, we completely neglect any inter-layer hopping for the planar $p_{x,y}$ derived states here. Correspondingly, the phase factor remains unchanged for the atoms within the layer, and so the B-AB splitting is unaltered from $\Gamma$ to A, and hence there is no $k_z$ dispersion. Although an oversimplification, this is a reasonable  approximation in the TMDs and other van der Waals layered materials where hopping strengths along the $c$-axis are naturally much larger for $p_z$ orbitals than $p_{x,y}$ orbitals. 

The net result is a series of band crossings of the $p$-orbital-derived states as a function of varying out-of-plane momentum. The crossing of the $p_z$-derived state with the top of each of the $p_{x,y}$-derived spin-orbit split pair remains protected. These crossings are between $R_4$ and $R_{5,6}$ bands, which give different eigenvalues under the operation of $C_{3v}$; their wavefunctions are therefore orthogonal and their crossing points can be described as lattice protected bulk Dirac points~\cite{yang_classification_2014, bahramy_ubiquitous_2018}. The other crossings are, however, gapped by spin-orbit coupling. Normally, one would consider such gaps to be topologically trivial, given that they both originate within $p$-orbital manifolds, and atomic $p$-orbitals are always of odd parity. However, there are two chalcogen atoms within the unit cell, located symmetrically about the transition metal which sits an the inversion centre (M-point), as described above. Bonding and anti-bonding states formed from these two $p$-orbitals enforce specific phase relations between the $p$-orbitals on the two sites, leading to states which can then have either even $(+)$ or odd $(-)$ parity. We show this explicitly for the $p_z$ orbitals in Fig.~\ref{f:mechanism}(c), where the bonding and anti-bonding combinations are of even parity, and odd parity, respectively. Hence a band parity inversion can readily be realised within a single orbital manifold as a function of band dispersion along the $k_z$-axis. We return to this point below.

The above discussion demonstrates that an array of inverted band gaps (IBGs) and BDPs can be produced with only two prerequisites. 
Firstly, there must be a disparity in  bandwidths along a direction in the Brillouin zone adhering to the rotational symmetry of the lattice,  which is naturally to be expected given anisotropy of orbital wavefunctions in all but $s$-orbitals. Secondly, this bandwidth disparity must be larger than the energetic separation of these bands imposed by the combination of the crystal field splitting (CFS) and the spin-orbit interaction. The combination of these two factors ensures multiple band crossings occur along a rotational axis, wherein bulk Dirac points, protected by the rotational lattice symmetry, and inverted band gaps can be simultaneously produced.

\begin{figure*} 
\includegraphics[width=\textwidth]{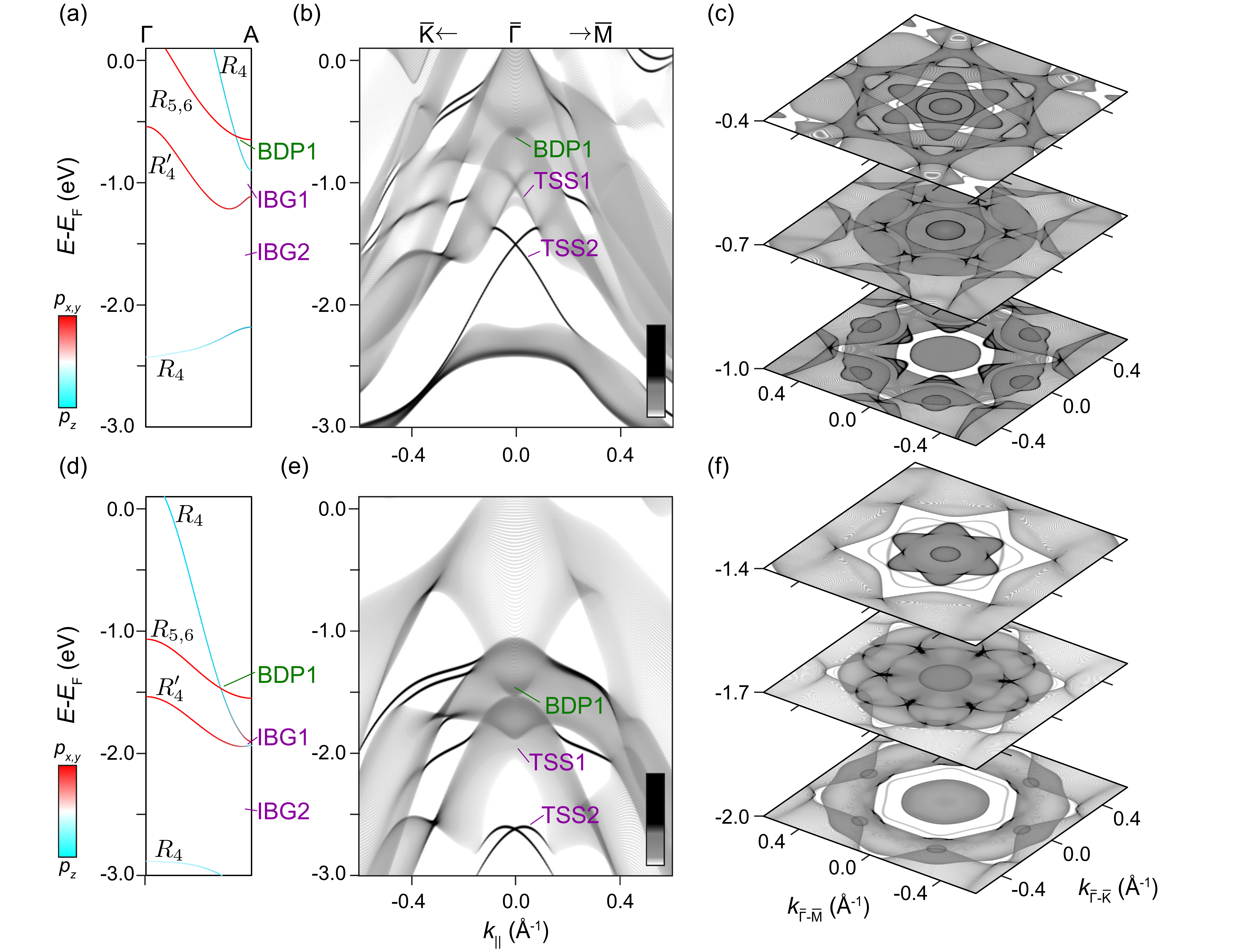}
\caption{Topological ladders and small $k_z$-projected band gaps in PtSe$_2$ and PdTe$_2$. (a) Orbitally projected $k_z$ DFT calculation along the $\Gamma$-A direction of PdTe$_2$.  bulk Dirac points (BDPs) and inverted band gaps (IBGs) are labelled (b) Corresponding surface slab calculation along the in-plane directions. Topological surface states, TSS1 and TSS2, are labelled. (c) Constant energy $k_x$-$k_y$ contours of PdTe$_2$ produced by surface slab calculations. (d-f) Equivalent plots for PtSe$_2$. }
\label{f:slabcalcs}
\end{figure*}

In a real system, the interlayer hopping between $p_{x,y}$ orbitals will not be zero and so $p_{x,y}$-derived bands will not be dispersionless along $k_z$. Nevertheless, this model is very well realised in 1T-PdTe$_2$ and 1T-PtSe$_2$, demonstrated in Fig.~\ref{f:slabcalcs} via density functional theory (DFT) based bulk and surface slab calculations. Despite the reduction of spin-orbit coupling from the telluride to the selenide, in each case a type-II bulk Dirac point and two parity inverted band gaps are formed below the Fermi level along the $\Gamma$-A line, with only the positioning and band gap size differing between the compounds. In each case, the type-II BDP is formed from the lattice protected crossing of AB-$R_4^-$ and B-$R_{5,6}^-$, where the superscript indicates the band parity. The highest binding energy IBG is formed from the anti-crossing of B-$R_4^+$ and B-$R_{4}'^-$, with the shallower IBG formed from an anti-crossing of B-$R_{4}'^-$ with both the B- and AB-$R_4$ bands of $+$ and $-$ parity respectively to create an overall non-trivial anticrossing~\cite{bahramy_ubiquitous_2018}.

In each of these compounds the topological ladder  centred at $k_{\parallel}$=0 is thus composed of a single type-II bulk Dirac cone and two topological surface states below $E_F$, vertically offset in energy.  These are labelled ``BDP1'' and ``TSS\{1,2\}'' in Fig.~\ref{f:slabcalcs}. One or more constituent states of these topological ladders have been discussed in other works~\cite{yan_identification_2015, huang_type_2016, zhang_experimental_2017, bahramy_ubiquitous_2018, clark_fermiology_2018, li_topological_2017}.

The calculations shown in Fig.~\ref{f:slabcalcs}, however, reveal a much richer surface electronic structure than this simple picture suggests. Although the $p_{x,y}$-derived bands are relatively non-dispersive when compared to the $p_z$-derived band, Fig.~\ref{f:slabcalcs}(a, d) shows how they in fact still have significant $k_z$ dispersion. Moreover, they each possess large bandwidths along the in-plane momentum directions, $k_{\parallel}$, owing to significant $p_{x,y}$ intra-layer hopping.

The surface slab calculations in Fig.~\ref{f:slabcalcs}(b-c, e-f) show how this inherent three-dimensionality of the chalcogen $p$-orbital manifold severely complicates the band structure away from $k_{\parallel}=0$, with multiple overlapping sets of states (Fig.~\ref{f:slabcalcs}(c, f)). In particular, projecting the bulk states onto the surface yields only small regions where true surface-projected band gaps are obtained. Topological surface states must necessarily thread through these small band gaps while connecting between neighbouring TRIM points~\cite{fu_topological_2007} (here $\Gamma$, A, L, M), and so these small projected band gaps greatly confine the paths that these surface states can traverse. The implications of this will be discussed below.

\section{Influence of small $k_z$-projected band gaps in the group-X TMDs}

\begin{figure*}
				\includegraphics[width=\textwidth]{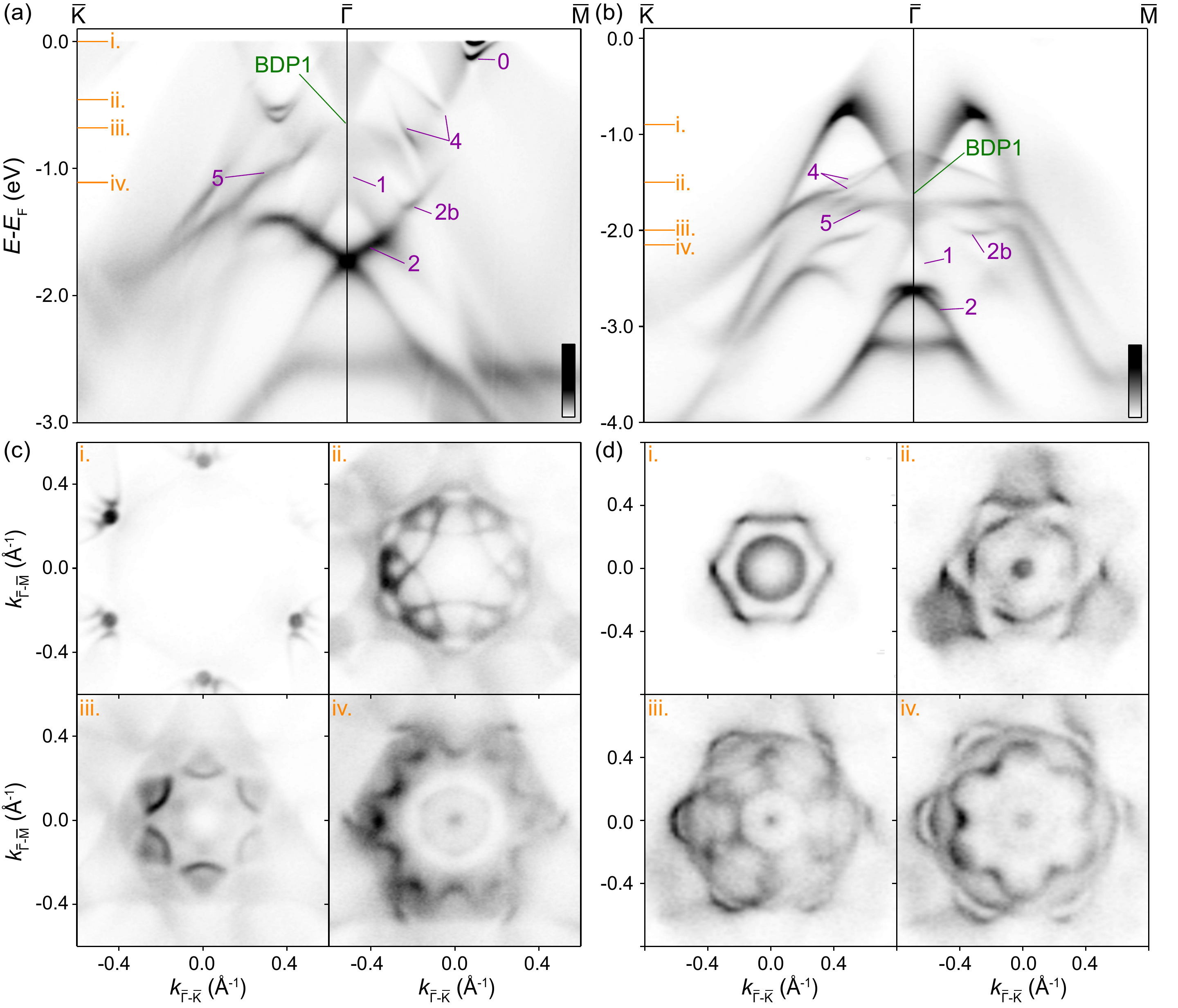}
				\caption{{Experimentally determined surface electronic structures of PdTe$_2$ and PtSe$_2$.} (a) Dispersions of PdTe$_2$ as obtained by ARPES (h$\nu=$24~eV, probing close to an A plane) along the $\overline{\text{K}}$-$\overline{\Gamma}$-$\overline{\text{M}}$ direction of the surface BZ. Surface states and the bulk Dirac point (BDP1) are labelled. (b) An equivalent dataset for PtSe$_2$ (h$\nu=$107~eV, probing close to an A plane). (c) Selected constant energy contours ($\pm$10~meV) of PdTe$_2$ (h$\nu=$27~eV) at the energies indicated in (a). (d) Selected constant energy contours ($\pm$10~meV) of PtSe$_2$ (h$\nu=$107~eV) at the energies indicated in (b).}
				\label{f:overview}
			\end{figure*}

\begin{figure*} 
			\includegraphics[width=\textwidth]{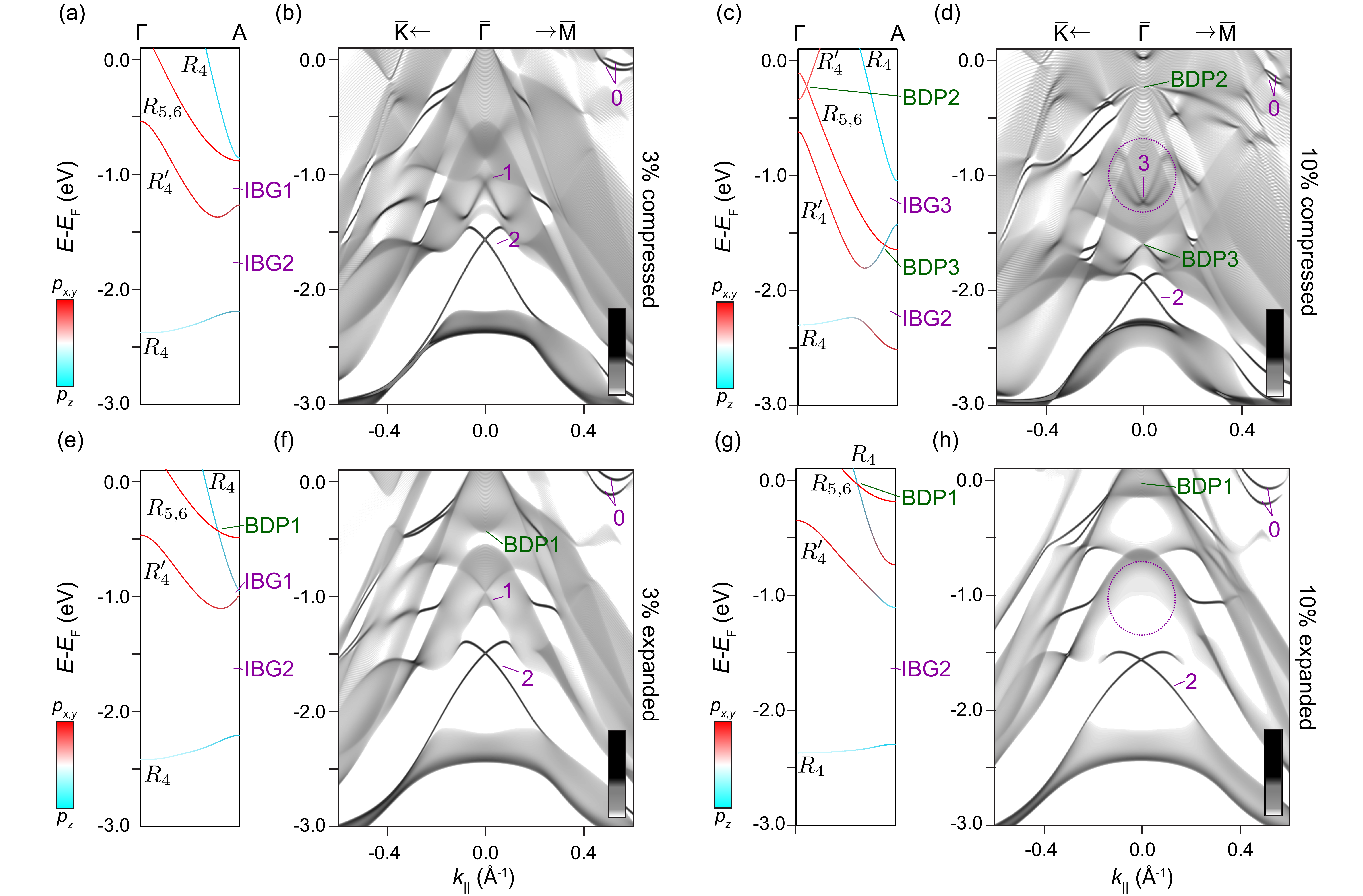}
				
				\caption{Creation and annihilation of bulk and surface Dirac fermions with uniaxial  strain. (a) Orbitally resolved bulk band structure of PdTe$_2$ along $\Gamma$-A direction, assuming a 3\% uniaxial compression along the $c$-axis. (b) The corresponding surface projection along $\overline{\text{K}}$-$\overline{\Gamma}$-$\overline{\text{M}}$ path. BDP1 is lost, but both inverted band gaps, hosting TSS1 and TSS2, remain. (c) and (d) show corresponding calculations for 10\% compressive strain. Now TSS1 is lost, while two new type-I BDPs are formed, as well as an additional IBG hosting TSS3. (e) and (f) The respective bulk and slab calculations with 3\% uniaxial expansion. All band inversions persist as in the unstrained case, but the IBG hosting TSS1 has nearly closed. (g) and (h) With 10\% expansion, now TSS1 is lost. This is due to an upward shift of bonding $R_{4}'$ ($p_{x,y}$) state, avoiding its crossing with bonding $R_{4}$ ($p_{x,y}$) state. The crossing with the upper (antibonding)$R_{4}$ state still results in a hybridised gap, but can not create a TSS, as they both share the same parity character.}
				\label{f:strain}
			\end{figure*}

\begin{figure*} 
				\includegraphics[width=\textwidth]{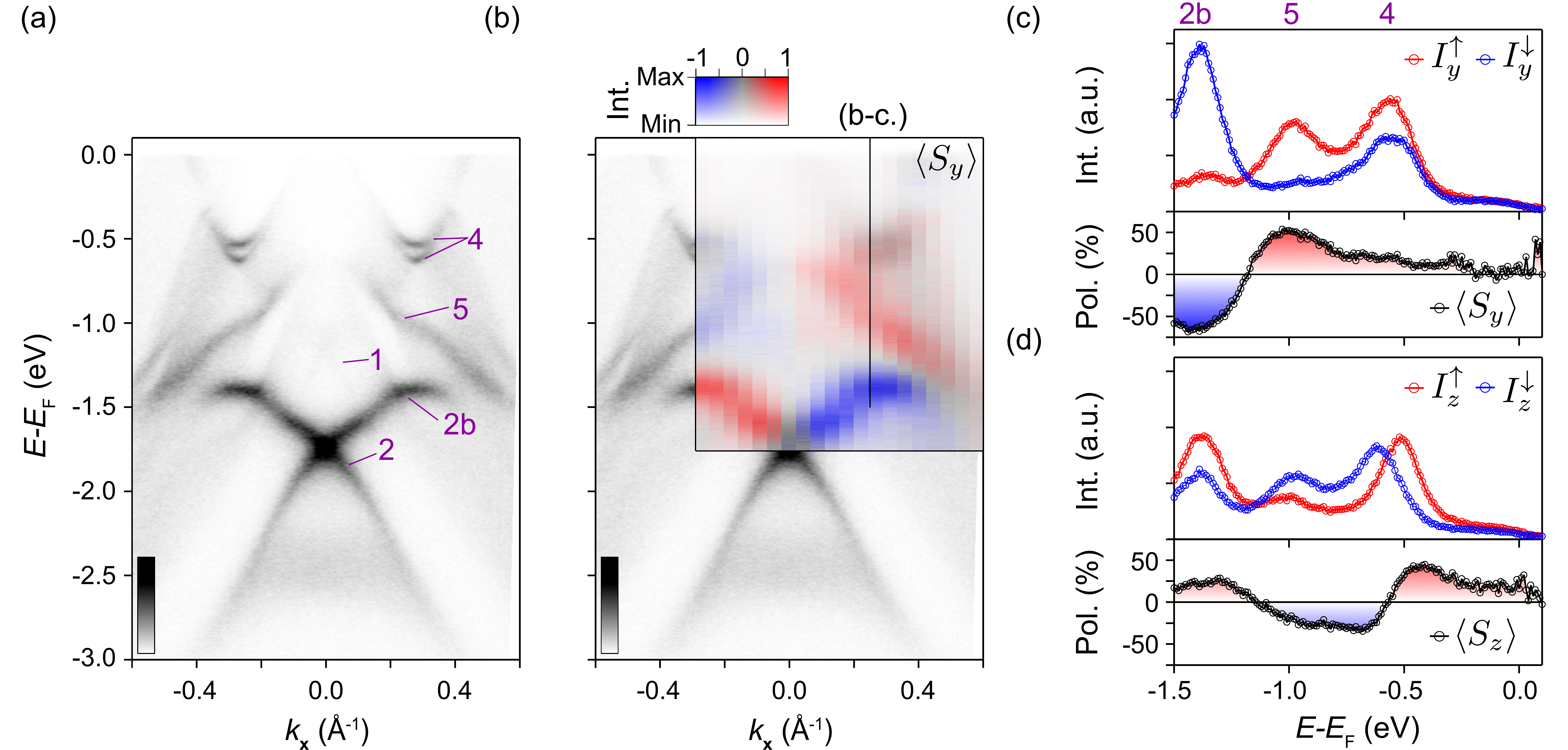}
				\caption{ Spin-polarisations of surface states 2b, 4 and 5 in PdTe$_2$ (a) ARPES dispersion along the $\overline{\Gamma}$-$\overline{\text{K}}$ direction (h$\nu=$29~eV). (b) The dispersion in (a) is duplicated, with a spin-resolved dispersion (h$\nu=$29~eV) overlaid for the chiral spin component ($\langle S_y \rangle$). This spin-resolved dispersion utilises a 2D colour scale (see Methods). (c-d) Spin-resolved constant energy-distribution curves (EDCs) (h$\nu=$27~eV) at the position indicated in (b) for the chiral ($\langle S_y \rangle$) (c) and out-of-plane ($\langle S_z \rangle$) spin components.}
				\label{f:PTSpin}
			\end{figure*}
			
Fig.~\ref{f:overview} provides an overview of the electronic band structures of PdTe$_2$ and PtSe$_2$ as measured by angle-resolved photoemission (ARPES) measurements performed using photon energies chosen to probe close to an A plane of the three-dimensional Brillouin zone. Surface states (the prefixes `(T)SS' are dropped), along with the type-II BDP, are indicated. BDP1, TSS1 and TSS2 form the `topological ladder' discussed above~\cite{huang_type_2016, bahramy_ubiquitous_2018, clark_fermiology_2018}.

Several other sharp states, which we attribute as additional surface states, are also visible within these band dispersions. The surface state labelled `0' here is a TSS resulting from a band inversion along the $k_z$ axis above the Fermi level, an analogue of the IBG that generates TSS2 below the Fermi level~\cite{clark_fermiology_2018}. This is present only in PdTe$_2$, while a smaller B-AB splitting in PtSe$_2$ renders this band inversion absent in the latter. The Dirac point of the resulting surface state in PdTe$_2$ (TSS0) is located approximately 1~eV above $E_F$~\cite{clark_fermiology_2018}. However, as for TSS2 in PtSe$_2$, the upper legs of the surface state turn over away from $k_{\parallel}=0$ to avoid becoming degenerate with the surrounding $k_z$-projected bulk manifold. For TSS0, both the upper and lower legs follow narrow $k_z$-projected band gaps down to below the Fermi level, becoming experimentally observable only approximately midway along the $\overline{\Gamma}$-$\overline{\text{M}}$ direction~\cite{clark_fermiology_2018}.  

Although TSS0 shares an identical origin to TSS1 and TSS2, its position at high $k_{\parallel}$ renders this state severely modified away from circular geometry, instead forming complex multi-valley pockets in the Fermi surface of PdTe$_2$, shown in Fig.~\ref{f:overview}(c i.). Although it retains an overall global chiral spin texture, it develops  significant radial spin canting as a result of this warping~\cite{clark_fermiology_2018}. TSS1 and TSS2 can be expected to become similarly increasingly warped with distance from $\overline{\Gamma}$, moulded by the surrounding $k_z$-projected band gaps.

Our measurements and calculations also reveal a number of additional surface states (SS) below $E_F$, visible at non-zero $k_{\parallel}$, and apparently common between PdTe$_2$ and PtSe$_2$. Firstly, from the spectral weight of TSS2, it is clear that a pronounced surface spectral weight is maintained even as the surface state becomes degenerate with the bulk manifold of states (the latter being evident as diffuse spectral weight). TSS2 therefore survives as a topological surface resonance, akin to the surface resonances discussed in the unoccupied states of Bi$_2$Se$_3$~\cite{jozwiak_spin_2016}. A similar phenomonology can be seen in our surface slab calculations shown in Fig.~\ref{f:slabcalcs}. At lower binding energies, the surface resonance emerges from the bulk continuum into a new projected band gap, forming a true surface state again which we assign as ``TSS2b'' in in Fig.~\ref{f:overview}, which can therefore be understood as a continuation of TSS2.  The relation between TSS2 and TSS2b is most obvious in PdTe$_2$, but a similar state can be assigned in PtSe$_2$, based in part on the spin-resolved measurements shown below. Two further surface states, labelled 4 and 5 in Fig.~\ref{f:overview}, are however seemingly distinct from the topological surface states contributing to the $\overline{\Gamma}$-centred topological ladder discussed above and in previous works. These further surface states are not {\it a priori} topological in origin, although we note that trivial surface states are uncommon in  van der Waals layered systems which cleave without `dangling bonds' or residual surface charge. 

Like TSS2b, both SS4 and SS5 exhibit rather unusual band  dispersions, moulded by the surrounding $k_z$-projected bulk manifold. For SS4 in PdTe$_2$, its two branches have a turning point mid-way along the $\overline{\Gamma}$-$\overline{\text{K}}$ direction, but monotonic downward dispersions along $\overline{\Gamma}$-$\overline{\text{M}}$.  Together, this produces a strongly hexagonaly warped, interlocking constant energy contour (Fig.~\ref{f:overview}(c ii.)), evolving to arc like-features when probing below the band minima along $\overline{\Gamma}$-$\overline{\text{K}}$ (Fig.~\ref{f:overview}(c iii)). This is not a Fermi arc of the form that connects Weyl points~\cite{lv_observation_2015, xu_observation_2015, xu_discovery_2015}. Rather, it represents how rich momentum-dependent surface state dispersions are shaped from the complex momentum-space structure of the bulk band gaps: the surface bands follow $k_z$-projected narrow channels which are inequivalent along $\overline{\Gamma}$-$\overline{\text{K}}$ and $\overline{\Gamma}$-$\overline{\text{M}}$, therefore producing apparently fragmented constant energy contours like the one seen here.

SS5 is a seemingly single-branch state that disperses up to, and passes through, the bulk Dirac point in both compounds.
In PdTe$_2$, the state makes an apparent crossing point with TSS2b partway along the $\overline{\Gamma}$-$\overline{\text{M}}$ direction. In contrast, along $\overline{\Gamma}$-$\overline{\text{K}}$ TSS2b appears to turn over, while the dispersion of SS5 also exhibits an abrupt change of slope, likely indicating a hybridisation-induced anti-crossing between the two states. The signatures of SS5 are more pronounced and its dispersion can be traced almost to $\overline{\text{K}}$.  
Fig.~\ref{f:overview}(c iv) shows the rich hybridised surface band structure that results, as evident from 
a constant energy contour of PdTe$_2$ for a binding energy between the band maximum of TSS2b along $\overline{\Gamma}$-$\overline{\text{K}}$, and the $\overline{\Gamma}$-$\overline{\text{M}}$ directions. This shows how TSS2b and SS5 together make an almost 12-fold symmetric pattern, highlighting that it is not only hybridisation of surface with bulk states that must be considered, but also hybridisation of various surface states themselves that can shape the hierarchy of the surface electronic structure of this system. 

Indeed, the complex and intertwined surface band network that results significantly complicates the isolation of the various surface states here. For example, since the band dispersion of SS5 is seemingly related to the position of the BDP, it is  tempting to assign this as a Fermi arc state persisting even when the bulk Dirac points from isolated $\pm{k_z}$ points are projected onto the same surface location, a point we return to below.

In PtSe$_2$, similar sets of surface states (2b, 4 and 5) can be assigned as in PdTe$_2$. They do, however, exhibit qualitative differences. Specifically, each has a much flatter band dispersion (Fig.~\ref{f:slabcalcs} and Fig.~\ref{f:overview}). We attribute this to details of the $k_z$-projected bulk manifolds, which in turn set the energy scales dominating the dispersion of the surface states through the projected band gaps. This, of course, depends on details of the $k_z$-dependent bandwidths, bonding-antibonding splittings due to differing transition metal electronegativities, and band overlaps, where the surface states disperse between different bulk manifolds. In fact, in PtSe$_2$, TSS2b becomes almost dispersionless along a narrow region of the $\overline{\Gamma}$-$\overline{\text{M}}$ direction of the Brillouin zone.

The rich and complex nature of these surface electronic structures, depending on fine details such as exact $k$-space locations of projected band gaps, precludes making a direct assignment of all features observed experimentally and those found in surface slab calculations. Nonetheless, both show the same key features (as evident from a comparison between Figs.~\ref{f:slabcalcs} and~\ref{f:overview}) and similarly complex surface electronic states dispersing between narrow bulk $k_z$-projected gaps. Moreover, additional insight can be gained by tuning the bulk electronic structure in the calculations by applying large uniaxial strains along the crystallographic $c$-axis (see Methods).

 Fig.~\ref{f:strain}(a-d) shows the case for PdTe$_2$ where a strain field is applied in compression, i.e. where the Pd-Te bond lengths remain unchanged but the interlayer gap decreases in size. The increased inter-layer hopping in this case enhances the size of the B-AB gap. This leads to a shifting of the AB-$p_z$ derived band to lower binding energies, and hence an opposite shifting of the B-$p_{x,y}$ derived bands occurs in order to maintain charge neutrality of the  system.  Accordingly, for the case of a 3\% strain (Fig.~\ref{f:strain}(a-b)), the type-II BDP (BDP1) formed from the crossing of AB-$R_{4}^-$ and B-$R_{5,6}^-$ is lost. By comparing Fig.~\ref{f:slabcalcs} and~\ref{f:strain}, it appears that only details in the band dispersions of the surface electronic structure changes here. With continued applied strain in compression (Fig.~\ref{f:strain}(c-d)), the picture changes substantially, however. At 10\% compression, the band inversion forming TSS1 is lost, but a new IBG as well as two additional BDPs are created, each centred at $\overline{\Gamma}$. The new IBG, positioned at $E-E_F \approx$~-1.2~eV, produces TSS3 in Fig.~\ref{f:strain}(d), formed from the anticrossing of the $p_z$-derived B-$R_{4}^+$ and AB-$R_{4}^-$ bands.
	
The type-I bulk Dirac point labelled BDP2 in Fig.~\ref{f:strain}(d) is formed from the crossing of AB-$R_{4}'^+$ and B-$R_{5,6}^-$, both of which have $p_{x,y}$ orbital character. BDP2 is located within 250~meV of $E_F$. BDP3, again of type-I character, is formed from the crossing of the $p_z$-derived B-$R_{4}^+$ with the $p_{x,y}$-derived B-$R_{5,6}^-$. We note that the formation of a type-I BDP with strain was presented previously in~\cite{xiao_manipulation_2017}, consistent with the results here. Crucially, new well-resolved surface states are visible dispersing through both BDP2 and BDP3. This supports our speculations above that SS5, observed clearly in experiment although not resolved in the slab calculations, could indeed be a surface state dispersing through the bulk Dirac point (BDP1) formed in the unstrained compound. 
		
Fig.~\ref{f:strain}(e-h) show equivalent calculations but with tensile strain. Here, the AB-$p_z$ derived band shifts to a higher binding energies with the B-$p_{x,y}$ derived bands moving to lower binding energies. Whilst AB-$R_{4}^-$ still anticrosses B-$R_{4}'^-$, the anticrossing between B-$R_{4}^+$ and B-$R_{4}'^-$ no longer occurs. There is therefore no longer a  parity exchange across the resultant band gap, and so TSS1 is lost. By comparing Fig.~\ref{f:slabcalcs} and~\ref{f:strain}, the overall surface electronic structure seems to be largely unaffected by this change, apart from the absence of TSS1 in Fig.~\ref{f:strain}(h). This again suggests a distinct origin of SS4 and SS5 to that of the topological states populating the topological ladder at $\overline{\Gamma}$ in the pristine compound. 
	
Naively, one might expect topological surface states to be lost with such large perturbations. This is clearly not the case here. All of the TSSs survive experimentally large strain values of 3\%, while TSS2, for example, survives even up to 10\% of both compression and expansion. This points to the extreme resilience of topological surface states formed from the mechanism introduced above of $k_z$-dependent band inversions occurring within a single orbital manifold. Even when the strain fields become unphysically large such that some topological gaps or bulk Dirac points are destroyed, new ones are naturally created in their place, again pointing to the ubiqiutous nature of the form of topological band inversions considered here. In fact, they even allow for forming Dirac states from crossings of the same orbital component (e.g. $p_{x,y}$ with $p_{x,y}$ as for BDP2 in Fig.~\ref{f:strain}(d).)
	
We return now to the topological surface states existing at zero strain, i.e., for the experimental crystal structures. We show below how the complexity of the surface state band dispersions evident here (see, e.g. Fig.~\ref{f:overview}(c,d)) in turn manifests via complex spin textures around the constant energy contours, commensurate with their highly-warped band contours as imposed by the small $k_z$-projected band gaps discussed above. 

Fig.~\ref{f:PTSpin} shows spin-integrated (a) and spin-resolved (b) $\overline{\Gamma}$-$\overline{\text{K}}$ dispersions of PdTe$_2$, with the surface states again indicated. The spin-dispersion shows a substantial `chiral' spin component (here $\langle S_y \rangle$), as expected for conventional topological and Rashba surface states~\cite{Rashba84}. This reveals how the clockwise chirality of TSS2 is inherited by TSS2b, with the spin-polarisation non-zero even over the region of $k_{\parallel}$ where the state is degenerate with the bulk manifold. This supports the above assignment of TSS2b as a continuation of TSS2. SS5 again has a significant CCW chirality, and switches its spin polarisation either side of the BDP. This spin texture is again consistent with this surface state forming a single spin-polarised state that disperses through the BDP.

SS4 is also strongly spin polarised, but unlike for the majority of the other surface states considered here, its chiral spin polarisation is rather small. Instead, spin-resolved energy distribution curves shown in Fig.~\ref{f:PTSpin}(c-d) demonstrate that it has a significant out-of-plane spin polarisation, of opposite sign for the two branches which turn over mid-way along the $\overline{\Gamma}-\overline{\text{K}}$ direction. We find that TSS2b and SS5 also host a finite out-of-plane spin component, however, this is much smaller than for the legs of SS4.

\begin{figure} 
				\includegraphics[width=\columnwidth]{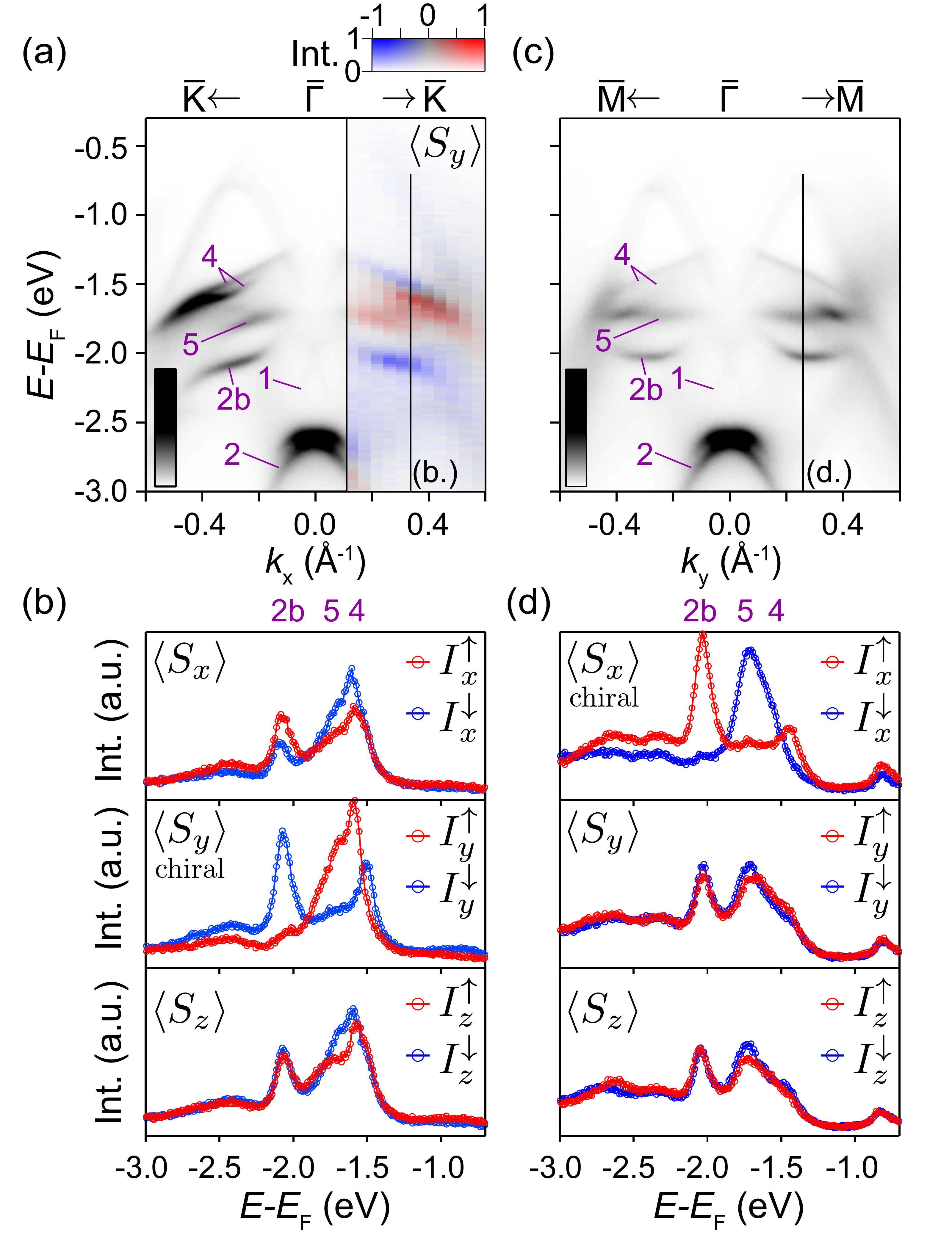}
				\caption{ Spin-polarisations of surface states 2b, 4 and 5 in PtSe$_2$ (a) ARPES dispersion along the $\overline{\Gamma}$-$\overline{\text{K}}$ direction (h$\nu=$29~eV) with a spin-resolved dispersion (h$\nu=$29~eV) overlaid for the chiral spin component ($\langle S_y \rangle$). This spin-resolved dispersion utilises a 2D colour scale (see Methods). (b) Spin-resolved EDCs (h$\nu=$29~eV) for the radial ($x$), chiral ($y$) and out-of-plane ($z$) spin components, for the position indicated in (a). (c) ARPES dispersion along the $\overline{\Gamma}$-$\overline{\text{M}}$ direction (h$\nu=$29~eV). (d) Spin-resolved EDCs (h$\nu=$29~eV) for the radial ($y$), chiral ($x$) and out-of-plane ($z$) spin components, for the position indicated in (b).}
				\label{f:PSSpin}
			\end{figure}

Fig.~\ref{f:PSSpin} details the spin textures of the corresponding surface states in PtSe$_2$. Each surface state retains the same sense of chirality $\langle S_y \rangle$ ($\langle S_x \rangle$) for Fig.~\ref{f:PSSpin}(a) ( Fig.~\ref{f:PSSpin}(b)) as in PdTe$_2$. However, in line with their monotonic dispersions, the chiral component is now dominant for all surface states, which can even be resolved for the two branches of SS4 which have opposite direction of chirality. These states do still exhibit a spin canting, however, with both a non-zero radial ($\langle S_x \rangle$ in Fig.~\ref{f:PSSpin}(a)) and out-of-plane ($\langle S_z \rangle$) components measured along the $\overline{\Gamma}$-$\overline{\text{K}}$ direction. Both of these components are symmetry enforced to reduce to zero along the $\overline{\Gamma}$-$\overline{\text{M}}$ direction, consistent with the experimental observations here. 

We note that the extensive warping of the surface state band dispersions, and corresponding spin texture evolution, observed here are somewhat reminiscent of the known warping of prototypical cononical topological surface states. There, it results as a natural consequence of higher order terms to the $\mathbf{k}\cdot\mathbf{p}$ Rashba Hamiltonian~\cite{fu_hexagonal_2009}, becoming increasingly pronounced at higher momenta. The warping of the surface states induced here, however, results predominantly due to shape of the underlying small $k_z$-projected band gaps. In fact, we show below how an inherent competition can even be found between these two warping mechanisms.

\begin{figure*} 
				\includegraphics[width=\textwidth]{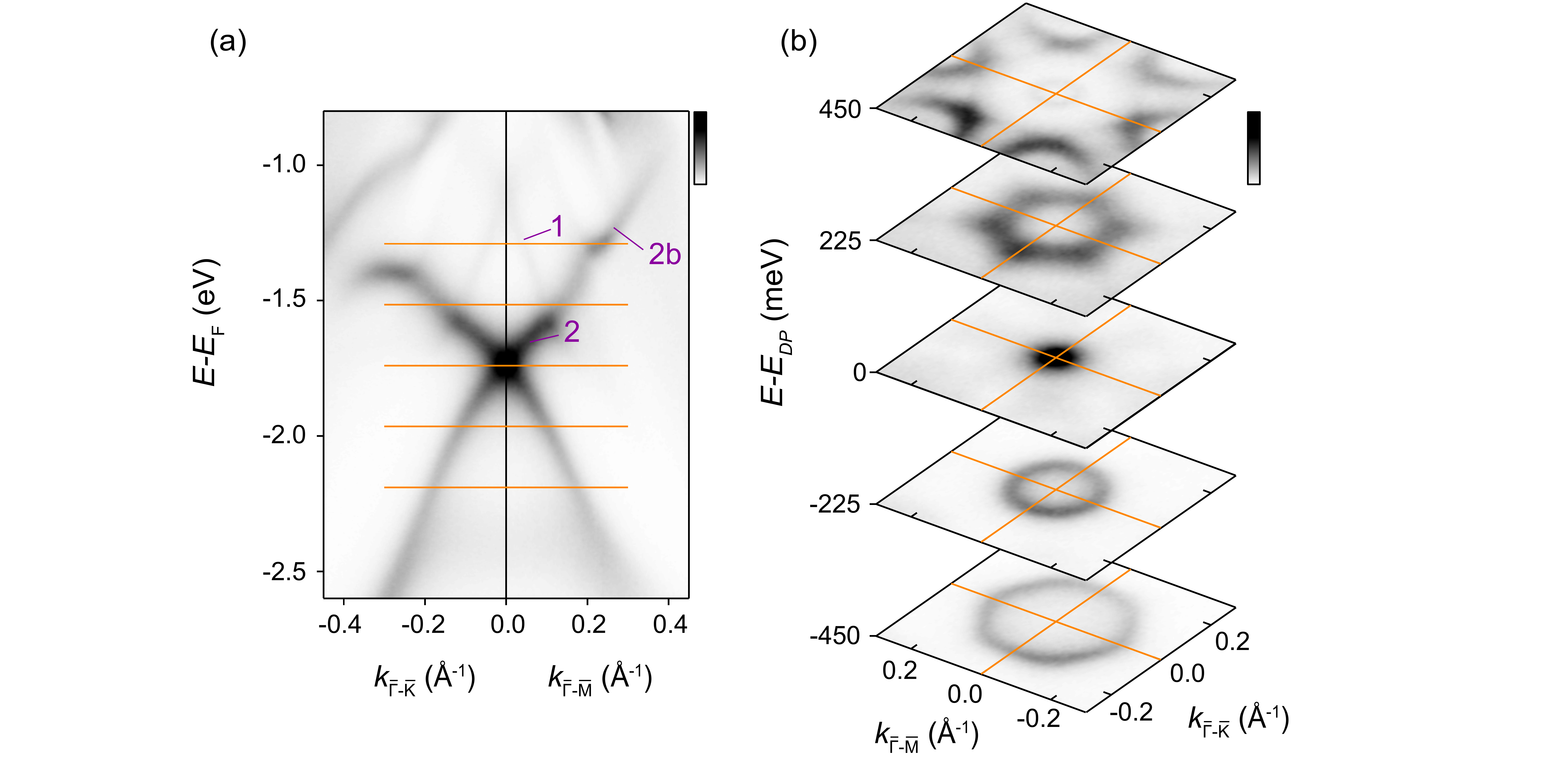}
				\caption{{Competition of hexagonal warping in PdTe$_2$} (a) ARPES dispersions ($h\nu=$27~eV) in the vicinity of TSS2. Orange lines correspond to the energy and momentum ranges of the constant energy contours in (b). (b) Constant energy contours ($\pm$15~meV) ($h\nu=$27~eV) for the energies indicated in (a). (c) Equivalent constant energy contours as determined by  DFT-based surface slab calculations.}
				\label{f:warping}
			\end{figure*}
			
To demonstrate this, we focus again on TSS2 in PdTe$_2$. Below its Dirac point, where the lower legs populate a large $k_z$-projected band gap, constant-energy contours shown in Fig.~\ref{f:warping} grow increasingly hexagonally warped with distance from the Dirac point, with hexagon corners orientated along the $\overline{\Gamma}$-$\overline{\text{M}}$ directions. This is entirely consistent with the $\mathbf{k}\cdot\mathbf{p}$ model of Fu~\cite{fu_hexagonal_2009}, which has been shown to be widely applicable for topological insulators and systems with Rashba surface states~\cite{fu_hexagonal_2009, frantzeskakis_anisotropy_2011, basek_spin_2011}. In contrast, above the Dirac point here, where the momentum-space extent of the $k_z$-projected band gap becomes smaller, the sign of the warping rotates by 30 degrees, with the apexes of the hexagonal states now formed along the $\overline{\Gamma}$-$\overline{\text{K}}$ direction.

This switch, incompatible with an intrinsic warping for a $C_{3}$ system, can again be explained in terms of the small $k_z$-projected band gaps. TSS2b has its warping direction set by the relative energetics of the local band gap of which it populates along the two azimuthal directions ($\overline{\Gamma}$-$\overline{\text{K}}$ or $\overline{\Gamma}$-$\overline{\text{M}}$). With reference to Fig.~\ref{f:overview}(a) and Fig.~\ref{f:warping}(a), TSS2b disperses into a band gap which is at a shallower binding energy when viewed along the $\overline{\Gamma}$-$\overline{\text{M}}$ direction than along the $\overline{\Gamma}$-$\overline{\text{K}}$ direction. Resultantly, the direction of warping of TSS2b ought to be such that its apexes are orientated along the $\overline{\Gamma}$-$\overline{\text{K}}$ direction, as verified by the shallowest energy constant energy contour in Fig.~\ref{f:warping}(b). TSS2 must follow suit to ensure good continuity to TSS2b. We expect that significant warping originating from small $k_z$-projected band gaps can be expected to be commonplace in compounds hosting $k_z$-mediated band inversions, given the small projected gaps in which these necessarily disperse.

The 1T-structured transition metal dichalcogenides were the first compound class shown to host topological ladders deriving from the mechanism outlined above. In addition to 1T-PdTe$_2$~\cite{bahramy_ubiquitous_2018, clark_fermiology_2018, yan_identification_2015, noh_experimental_2017} and 1T-PtSe$_2$~\cite{huang_type_2016, zhang_experimental_2017, bahramy_ubiquitous_2018, clark_fermiology_2018, li_topological_2017} discussed here, $k_z$ mediated topological ladders have been observed through a combination of angle-resolved photoemission and density functional theory in 1T-PtTe$_2$~\cite{yan_lorentz_2017},  1T-NiTe$_2$~\cite{xu_topological_2018} and 1T-IrTe$_2$~\cite{bahramy_ubiquitous_2018, fei_approaching_2017}.

\section{Universality of $k_z$ mediated band inversions}

This physics is not limited to the TMDs, however, and we stress that the prerequisites to realise this form of band inversion along a high-symmetry line is rather minimal. Indeed, it requires only crossings of dispersing bands along one or more rotationally symmetric axis, which can be expected to be commonplace. To this end, we show below that several recent realisations of bulk Dirac cones and topological surface states can be understood as consequences of the same underlying mechanism driving the formation of topological ladders in the TMDs, while additional examples can be found quite commonly in calculated electronic structures where they are yet to be identified explicitly. 

\begin{figure} 
			\includegraphics[width=\columnwidth]{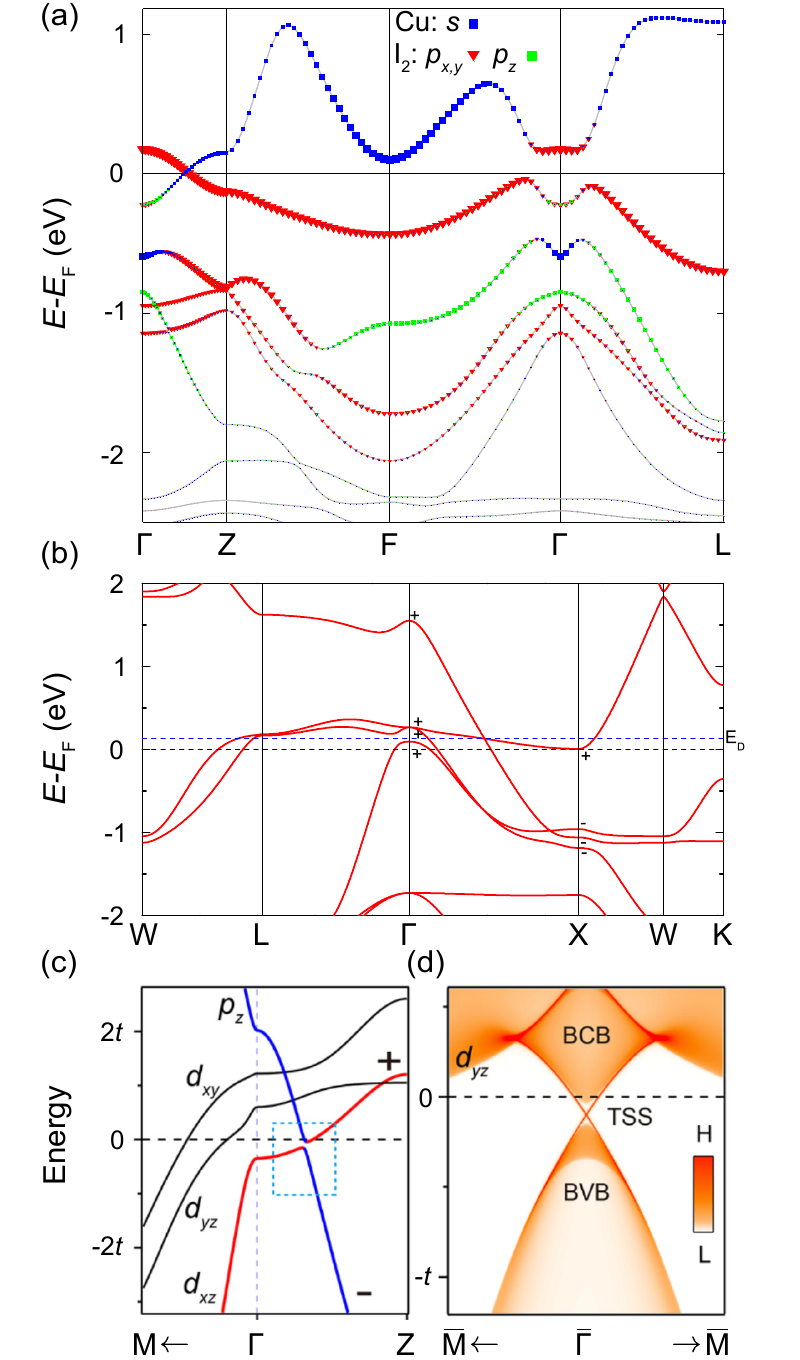}
				\caption{$k_z$-mediated topological phenomena outside of the TMD family. (a) Orbitally projected band structure calculation of $\beta$-CuI ($D_{3d}$) adapted from~\cite{le_dirac_2018}. (b) Band structure calculations of YPd$_2$Sn ($O_h$) reprinted with permission from~\cite{guo_type_2017}. Copyright (2019) by the American Physical Society. (c-d) Bulk (c) and surface (d) band structure calculations  FeSe$_{0.5}$Te$_{0.5}$ ($D_{4h}$)  from~\cite{zhang_observation_2018}, reprinted with permission from AAAS.}
				\label{f:others}
			\end{figure}

For example, a recent DFT study into the bulk band structure of $\beta$-CuI ($D_{3d}$) demonstrated the presence of a parity inverted band gap formed at $\Gamma$, as well as a type-I BDP at the Fermi level, formed along the $\Gamma$-Z direction~\cite{le_dirac_2018}.  Calculations of this system (Fig.~\ref{f:others}(a), \cite{le_dirac_2018}) also reveal a second type-I BDP and an anti-crossing gap formed along this same axis approximately 1~eV below $E_F$. This latter pair of crossings can be understood as a consequence of the mechanism shown in Fig.~\ref{f:mechanism}: The crystal field lifts the degeneracy of the iodine $p$-orbital manifold. The inclusion of spin-orbit coupling lifts the remaining degeneracy of the $p_{x,y}$-derived states (seen explicitly in further band structure calculations in~\cite{le_dirac_2018}), and a natural bandwidth disparity along the $C_3$-symmetric $\Gamma$-Z direction of iodine-derived $p$-bands produces the symmetry protected crossing and a spin-orbit mediated hybridisation gap, exactly as for the mechanism introduced above for the TMDs. 

Furthermore, the presence of trigonal symmetry is not a pre-requisite. The `Heusler' alloys ($O_h$) (\{Y, Sc,\}Pd$_2$Sn), (\{Zr, Hf,\}Pd$_2$Al) and (\{Zr, Hf,\}Ni$_2$Al), have been recently verified to host type-II bulk Dirac cones at the Fermi level~\cite{guo_type_2017}. At higher energies (approximately 1~eV below the Fermi level in YPd$_2$Sn), however, calculations (Fig.~\ref{f:others}(b)) again reveal an additional topological ladder, in this case forming an almost maximally-tilted type-I bulk Dirac cone and a gapped band crossing which is now formed along the $\Gamma$-X direction of the rock-salt Brillouin zone. Again, the $p_{x,y}$-derived bands are degenerate in the absence of spin-orbit coupling~\cite{guo_type_2017}. This therefore appears to be a cubic analogue of the physics introduced above. We note that parity inverted band gaps cannot form entirely within the $p$-orbital manifold of a rock-salt structured system, and so the gapped crossing is not an inverted band gap here. However, excitingly, six well-separated bulk Dirac points should be realised for this crystal structure. The $(001)$ surface should thus host giant Fermi arcs bridging between the surface projections of two of the three-pairs of bulk Dirac points.

We also note that the model presented here is not valid only for $p$-orbitals. An example is the 2H-structured TMDs, where similar topological ladders as for the 1T structured group-X systems are obtained, but with significant $d_{xz,yz}$-orbital character mixing into the $p_{x,y}$-character bands~\cite{bahramy_ubiquitous_2018}. Analogous physics can also be realised for systems with near pure $d$-orbital character of the relevant bands. An interesting example is the recent observation of bulk Dirac points and topological surface states in the Fe-based superconductors~\cite{zhang_multiple_2018, zhang_observation_2018}, which we argue form from an exactly analogous mechanism to the one presented here. In such systems, a $p_z$-derived band of either Se/Te (for FeSe$_x$Te$_{1-x}$) or As character (for LiFeAs, LaOFeAs, BaFe$_2$As$_2$) is always sufficiently dispersive to cross through the entirety of the Fe-derived $t_{2g}$ $d$-orbital manifold. As shown in Fig.~\ref{f:others}(c-d), a pair of bulk Dirac points are formed with the crossing of the $p_z$-derived band with the predominantly $d_{xy}$ and $d_{yz}$-derived bands, whilst the crossing with the band of predominantly $d_{xz}$ character becomes hybridised to produce a parity inverted band gap. We note that the resultant TSS in FeSe$_{0.5}$Te$_{0.5}$ crosses the Fermi level (Fig.~\ref{f:others}(d)). This has raised speculation that this compound could even be host to topological surface superconductivity~\cite{zhang_observation_2018, wang_evidence_2018}, although we note that in the model compound PdTe$_2$ introduced above, where TSS0 crosses the Fermi level (Fig.~\ref{f:overview}(a)), the superconductivity at the surface appears to be entirely conventional~\cite{clark_fermiology_2018, teknowijoyo_nodeless_2018}. 

We note that the general nature, and robustness, of the mechanism introduced here provides a powerful opportunity to study the interplay of $k_z$ mediated topological phenomena with many other bulk properties, and even their evolution when global lattice symmetries are lifted. 
As an example, the evolution of  bulk Dirac points, topological surface states and Fermi arcs with time-reversal (TRS) and inversion symmetry (IS) breaking may be possible, with potentially interesting consequences. For example, a theoretical study into the inversion asymmetric group-X Janus TMDs (PdSeTe, PtSSe, PtSeTe) shows how the bulk Dirac points formed along the $\Gamma$-A direction in (Pd,Pt)(Se,Te)$_2$ are transformed into pairs of three-fold degenerate `triple-points', midway between the two-fold and four-fold degeneracies of Dirac and Weyl points respectfully~\cite{bahramy_ubiquitous_2018, xiao_inversion_2018}.

The study of their evolution with time-reversal symmetry breaking may be possible in compounds with ferromagnetic ground states. For example, the transition metal carbides, \{Nb, Ta, V, Cr\}C ($O_h$), have been recently verified to host $k$-mediated bulk Dirac cones within their metal $d$-orbital manifolds~\cite{zhan_topologically_2018}. Whilst this compound series is non-magnetic, analogous physics can be expected to occur in the similarly structured transition metal nitride compounds, \{Sc, Ti, V, Cr, Zr, Nb\}N. These share remarkably similar band structures to the carbides~\cite{ma_first_2012}, but CrN has a ferromagnetic ground state. This could be used to realize intrinsic `topological magnetoelectric effects', wherein the Kramer's degeneracy of topological surface states is lifted~\cite{luo_massive_2013}, or novel Weyl or triple points stabilized in time-reversal asymmetric environments.

\section{Outlook}

The robustness of bulk Dirac points and topological surface states across many material families with disparate ground state properties and space group symmetries, coupled with the extremely minimal set of prerequisites to realise them, strongly suggests that topological ladders mediated by band inversions occurring along high-symmetry lines are a commonly-occurring feature within the electronic band structure of solids. Moreover, this same robustness allows broad possibilities for tuning these states. By altering interlayer hopping strengths~\cite{bahramy_ubiquitous_2018} through, for example strain~\cite{xiao_manipulation_2017, huang_type_2016} or by chemical substitutions~\cite{zhang_observation_2018, fei_approaching_2017}, the number, types, and energetic positions of of Dirac cones can be altered. This is in contrast to the much more limited control possible in conventional topological insulators. For example  modest changes of the spin-orbit coupling strength in the Bi$_2$Se$_3$ family removes the state of interest~\cite{zhang_topological_2009} and applying pressure to BiTeI can be used only as a switch for a topological surface state~\cite{bahramy_emergence_2012}.

More generally, the large array of compounds found to host topological phenomena via this new mechanism is in-line with the recent prediction that one quarter of all solids that exist in nature have within them a topological band inversion~\cite{vergniory_high_2018}. 
With time, it becomes increasingly clear that topological phenomena is prevalent in nature.

\

\begin{acknowledgements}
\noindent {\it Acknowledgements:} 
We gratefully acknowledge support from the Leverhulme Trust (Grant No.~RL-2016-006), the Royal Society, CREST, JST (Nos. JPMJCR16F1 and JPMJCR16F2), and the International Max-Planck Partnership for Measurement and Observation at the Quantum Limit. OJC acknowledges EPSRC for PhD studentship support through grant No. EP/K503162/1. IM acknowledges PhD studentship support from the IMPRS for the Chemistry and Physics of Quantum Materials. B.-J.Y. was supported by the Institute for Basic Science in Korea (Grant No. IBS-R009-D1) and Basic Science Research Program through the National Research Foundation of Korea (NRF) (Grant No. 0426-20170012, No.0426-20180011), the POSCO Science Fellowship of POSCO TJ Park Foundation (No.0426-20180002), and the U.S. Army Research Office under Grant Number W911NF-18-1-0137. This work was performed under the approval of Proposal Assessing Committee of HiSOR (Proposal No. 17BG022). We thank Diamond Light Source (via Proposal Nos. SI14927-1 and SI16262-1), HiSOR and Elettra Sincrotrone Trieste for access to the I05, BL9A and APE beamlines respectfully that contributed to the results presented here. 
\end{acknowledgements}

This is the Accepted Manuscript version of an article accepted for publication in Electronic Structure. 
IOP Publishing Ltd is not responsible for any errors or omissions in this version of the manuscript or any version derived from it. The Version of Record is available online at https://doi.org/10.1088/2516-1075/ab09b7.

\end{document}